\documentclass[smallextended,natbib,runningheads]{svjour3}
\journalname{Annals of the Institute of Statistical Mathematics}
\smartqed  
\usepackage{graphicx}
\usepackage{amsmath,amssymb}
\newtheorem{assumption}{Assumption}
\begin{document}

\title{Local Consistency of Markov Chain Monte Carlo Methods
\thanks{Supported in part by Grant-in-Aid for JSPS Fellows (19-3140)
and  Grant-in-Aid for Young Scientists (B) 22740055.
}
}


\author{Kengo KAMATANI
}


\institute{K. KAMATANI \at
             Graduate School of Engineering Science, Osaka University, 
Machikaneyama-cho 1-3, Toyonaka-si, Osaka, 560-0043 , Japan\\
              \email{kamatani@sigmath.es.osaka-u.ac.jp} }

\date{Received: date / Revised: date}

\maketitle

\begin{abstract}
In this paper, we introduce the notion of efficiency (consistency)
and examine some asymptotic properties of Markov chain Monte Carlo  methods. 
We apply these results
to the data augmentation (DA) procedure for independent and identically distributed observations. 
More precisely, we show that if both the sample size and the running time of the DA procedure
tend to infinity
the empirical distribution of the DA procedure tends to the posterior distribution.
This is a local property of the DA procedure, which may be, in some cases, 
more helpful than the global properties to describe its behavior. 
The advantages of using the local properties
are the simplicity and the generality of the results. 
The local properties provide
useful insight into the problem of how to construct efficient algorithms.
\keywords{Monte Carlo \and Markov chain \and Asymptotic Normality}
\end{abstract}

\section{Introduction}
This paper investigates conditions under which a Markov chain Monte Carlo (MCMC) procedure
has a good stability property in the Bayesian context.
There have a vast literature related to the sufficient conditions for ergodicity:
see reviews \cite{TierneyAOS94} and \cite{RR}
and textbooks such as \cite{N} and \cite{MT}. 
The transition kernel of the MCMC  procedure is Harris recurrent 
under fairly general assumptions. Moreover, 
it is sometimes geometrically ergodic. 
In practice, 
the Foster-Lyapunov type drift condition is commonly used
to establish geometric ergodicity and calculation of its rate.
This condition is helpful for studying the global property of the MCMC procedure. However there are some limitations 
if we want more information for the stability of the MCMC procedure; for example this approach has difficulty in comparing two MCMC procedures. 


We take another approach to study the stability of 
the MCMC procedure in the Bayesian context. 
We will define local consistency as a measure of the performance
of the  MCMC procedure. 
The following toy example illustrates our approach. 

\begin{figure}[htbp]
\includegraphics[width=8cm,bb=0 0 576 576]{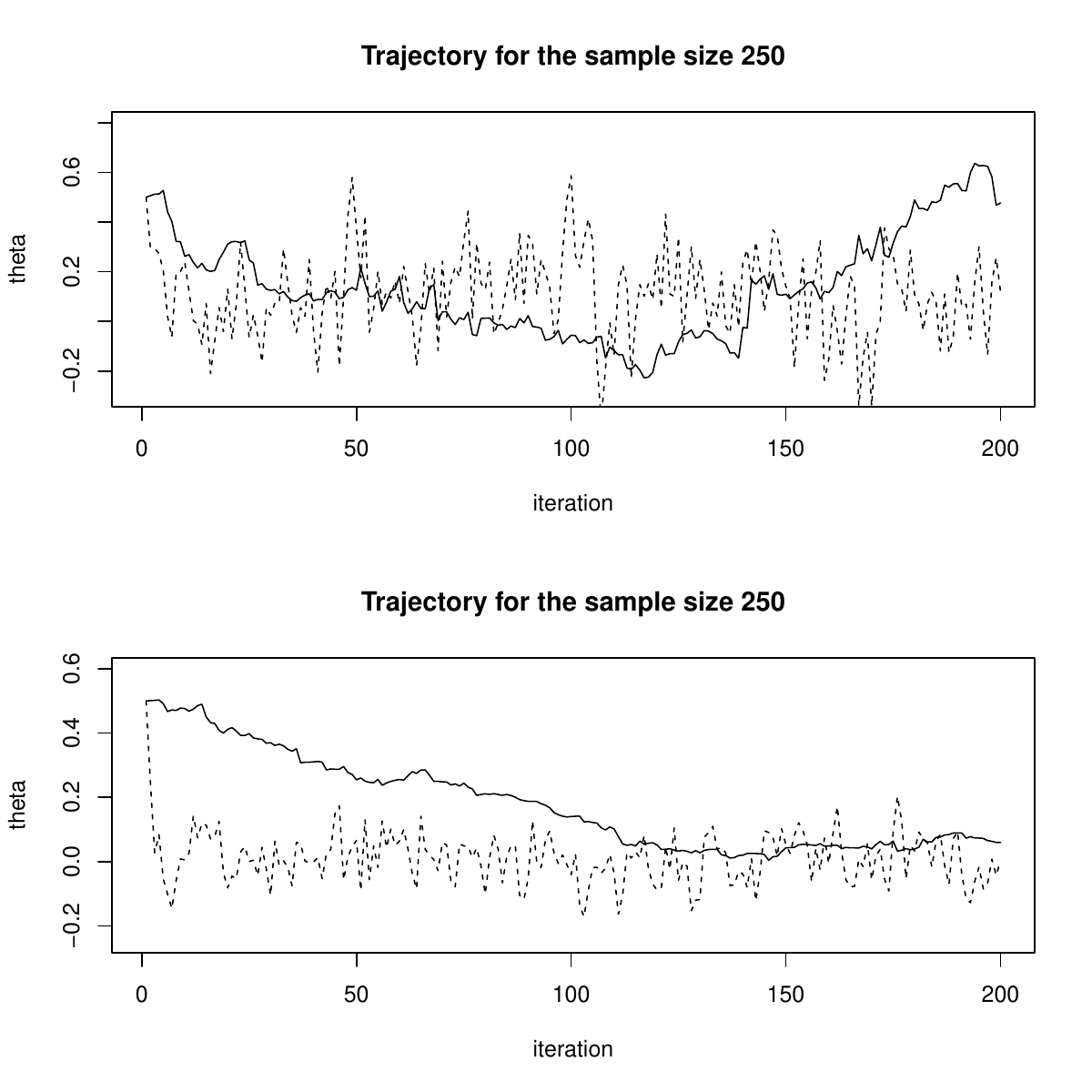}
\caption{Trajectory of the DA procedures for sample size $n=50$ (upper) and $n=250$ (lower). Solid line is for (\ref{eq2}) and dashed lines is for (\ref{eq3}).\label{Figure1}}
 \end{figure}

Assume we have $n$ observation $x_n=\{x^1,\ldots, x^n\}$ from a simple model
\begin{displaymath}
P(X=1|\theta)=\Phi(\theta),\ 
P(X=0|\theta)=1-P(X=1|\theta)
\end{displaymath}
where 
$\theta$ is the parameter
and $\Phi$ is the cumulative distribution function of the normal distribution. The data augmentation (DA) procedures can be defined by the so-called augmented data model that introduces latent variable
$y$. We consider two DA procedures corresponding to the following  augmented data models; 
\begin{eqnarray}
 y&\sim  N(0,1),\ x =&1_{\{y\le\theta\}}\label{eq2}\\
 y&\sim  N(-\theta,1),\ x =&1_{\{y\le 0\}}\label{eq3}.
 \end{eqnarray}
Though the models are similar, 
the performances of the DA procedures are quite different. 
Figure \ref{Figure1} is a trajectory of the sequences
\begin{displaymath}
\theta_0,\ldots, \theta_{m-1}
\end{displaymath}
from the DA procedures with the sample sizes $n=50$ and $n=250$. 
The true value is set to $\theta_0=0$. 

The simulation result for the sample size $n=50$ (upper)  shows the poor performance of the DA procedure for  (\ref{eq2}) than that for (\ref{eq3}) that may cause the inference bias. 
Such differences become clear (see $n=250$ (lower)) and the analysis
becomes easier as the sample size grows. 

With this observation in mind, we want to define consistency as an asymptotic property
as the sample size $n\rightarrow\infty$. 
For each observation 
$x_n$, the MCMC procedure
results in a Markov chain $\theta_\infty=\{\theta_0,\theta_1,\ldots\}$
that has the invariant probability distribution $p(d\theta|x_n)$. 
Write $\mathbb{P}_n$ for the probability measure for $x_n$ and $\theta_\infty$. 
Let
\begin{equation}
I=\int\varphi(\theta)p(d\theta|x_n),\ 
I_m=m^{-1}\sum_{i=0}^{m-1}\varphi(\theta_i)
\end{equation}
for a bounded continuous function $\varphi$.
The goal of the MCMC procedure is to approximate $I$ by $I_m$. 
It would be  helpful if 
\begin{equation}\label{eq0}
|I-I_{m_n}|=o_{\mathbb{P}_n}(1)
\end{equation}
for any $m_n\rightarrow\infty$. Since the number $m$ corresponds to the iteration counts of the MCMC procedure,  smaller is better. 
For each $x_n$ usually 
$\lim_{m\rightarrow\infty}|I-I_{m}|=0$ in $\mathbb{P}(\cdot|x_n)$ holds 
by ergodicity of the Markov chain. However sometimes the relation (\ref{eq0}) 
does not hold. 
For example, under a reasonable scaling, the DA procedure that uses the augmented data model (\ref{eq2})
does not satisfy (\ref{eq0}) for some $m_n\rightarrow\infty$
but it  does satisfy  if  $m_n/n\rightarrow\infty$. 
If (\ref{eq0}) holds for  any $m_n\rightarrow\infty$ we call
the MCMC procedures consistent. 
The DA procedure that uses
(\ref{eq3}) has consistency under the same scaling that will be proved in Theorem \ref{regugibbs}. 

We obtain the following results. 
\begin{enumerate}
\item
The consistency and the local consistency of the MCMC procedures are studied. 
\item 
A reasonable set of sufficient conditions for the local consistency for the DA procedure is addressed 
for independent identically distributed observations.
We only assume 
(a) the identifiability of parameter, (b) the existence of uniformly consistent test, (c)
regularity of prior distribution, and (d) quadratic mean differentiability of 
the full model. 
\end{enumerate}


For a treatment of a large sample setting (with a different motivation), 
a recent paper \cite{BC} studied 
the Metropolis algorithm for increased parameter dimension $d$. 
They obtained the rate of the running time of the Metropolis algorithm
for burn-in and after burn-in. To deal with the complex algorithm
and to obtain strong results, 
they assumed  strong conditions (C.1, C.2 and (3.5)).
Another paper, \cite{Nelson00} and \cite{Svenson09} obtained stability properties of the stochastic EM algorithm. 
Essentially they studied finite dimensional convergence of $\theta_0,\ldots, \theta_k$.
 However, without tightness arguments, the finite dimensional properties
are insufficient to describe the performance of the MCMC procedures.
On the other hand, we show the convergence of the law of the process $\{\theta_i;i\in\mathbf{N}_0\}$ with a minimal set of conditions. 

It is not our intention to conclude that the DA procedure is always efficient.
The conclusion of Theorem \ref{regugibbs} is that under regularity conditions
 the DA procedure approximates the posterior distribution
in an ordinal running time.  
On the other hand, it illustrates the causes of the performance bottlenecks  of the MCMC procedures.
For example, (a) the choice of the initial guess $\theta_0$ is not good, 
(b) the model has the fragility of the identification (c) the Fisher information matrix $g$ for the model is too small or that for the hidden information  is too large, or 
(d) the sample size is too small related to its parameter dimension. 
For example, the DA procedure that uses the model (\ref{eq2})
suffers from (c). 
These studies of regular/non-regular properties are quite important 
 for the elimination of the performance bottlenecks of the MCMC procedures.

The rest of the paper is organized as follows. 
We prepare in Section \ref{snrmc} for needed backgrounds. 
Consistency will be introduced in Section \ref{rmcp}. We analyze local consistency of the DA procedure in this section. 
Concluding remarks are summarized in Section \ref{cr}.


%
%

\section{Background}\label{snrmc}

\subsection{Quadratic mean differentiability}\label{qmd}
Let $(X,\mathcal{X})$ be a measurable space. Let $M=\{P_\theta(dx)=p_\theta(x)dx;\theta\in\mathbf{R}^d\}$
be a parametric family on $X$.
The  family $M$ is said to be quadratic mean differentiable at $\theta$ if
\begin{equation}
\sqrt{p_{\theta+h}(x)}-\sqrt{p_\theta(x)}-h'\tilde{\eta}=o(h)\ \mathrm{in}\ L^2(dx)
\end{equation}
for any $h\rightarrow 0$ and  a square integrable function $\tilde{\eta}:X\rightarrow\mathbf{R}^d$
where $v'$ is the transpose of a vector $v\in\mathbf{R}^d$. 
A matrix $g(\theta)=4\int \tilde{\eta}\tilde{\eta}'dx$ is called the Fisher information matrix. In this paper, when $x_n=\{x^1,\ldots, x^n\}\sim P_{\theta}^{\otimes n}=\prod_{i=1}^n P_\theta(dx^i)$, 
the following random variable is called the normalized score function:
\begin{equation}
\eta=\eta_\theta(x_n)=\sqrt{n}^{-1}\sum_{i=1}^n 2\frac{\tilde{\eta}(x^i)}{\sqrt{p_\theta(x^i)}}.
\end{equation}
Suppose now that $M^\dagger=\{P^\dagger_\theta(dxdy)=p^\dagger_\theta(xy)dxdy;\theta\in\mathbf{R}^d\}$
is another model, called the augmented data model on $X\times Y$ that satisfies 
$P_\theta(dx)=\int_Y P^\dagger_\theta(dxdy)$. 
According to  Proposition 7.4 of \cite{Cam1988}, 
if $M^\dagger$ satisfies quadratic mean differentiability at $\theta$ then $M$ also does.  
Let $g$ and $g^\dagger$ be the Fisher information matrices of the models 
$M$ and $M^\dagger$
and write 
$\eta$ and $\eta^\dagger$ for the normalized score statistics, and
write $\hat{u}$ and $\hat{u}^\dagger$ for the maximum likelihood estimators
of the models $M$ and $M^\dagger$
under observations $x_n$ and $x_n, y_n=\{y^1,\ldots, y^n\}$ with respectively.

Let $(\Omega,\mathcal{F},\mathbb{P})$ be a probability space and let $\mathcal{G}$ be a sub $\sigma$ algebra of $\mathcal{F}$. 

\begin{definition}[Stable convergence]
A sequence of $\mathbf{R}^d$-valued $\mathcal{F}$-measurable random variable $Z^n$ is said to converge $\mathcal{G}$-stably if there 
exists a measure $\mu$ on $\Omega\times\mathbf{R}^d$ such that
\begin{equation}
\mathbb{E}[f(Z^n)Y]\rightarrow\int \mu(d\omega,dx)Y(\omega)f(x)
\end{equation}
for any continuous bounded function $f$ and for any bounded $\mathcal{G}$-measurable random variable $Y$. 
\end{definition}

Let $X_i(\omega), Y_i(\omega)\ (i=1,\ldots, n)$
be i.i.d. observation from a probability measure $P_\theta^\dagger(dxdy)$
and set $X(\omega)=\{X_i(\omega);i=1,\ldots\}$ and $\mathcal{G}=\sigma(X)$. 

\begin{lemma}\label{stableconv}
Let $\eta^*=\eta^\dagger-\eta$ and $g^*=g^\dagger-g$. Then $\eta^*$ converges $\mathcal{G}$-stably to $N(0, g^*)$. 
\end{lemma}

\begin{proof}
By the law of large number, for almost all $\omega\in\Omega$, 
\begin{equation}
\mathbb{E}[\eta_i^*|\mathcal{G}]=0, \mathbb{E}[n^{-1}\sum_{i=1}^n\eta^*_i(\eta^*_i)'|\mathcal{G}]\rightarrow g^*
\end{equation}
and 
\begin{equation}
\mathbb{E}[n^{-1}\sum_{i=1}^n|\eta^*_i|^21_{\{n^{-1/2}|\eta^*_i|>\epsilon\}}|\mathcal{G}]\rightarrow 0
\end{equation}
for $\epsilon>0$. Write $A_\epsilon\in\mathcal{G}$ for all $\omega\in\Omega$ that satisfies the above three convergences. 
Then $A=\cap_{i=1}^\infty A_{i^{-1}}$ is still a sure event, and for each of 
$\omega\in A$,  the Lindeberg condition holds for 
the array $\{\eta_i^*(x_i, Y_i(\omega));i=1,\ldots, \}$ with probability measure $\mathbb{P}(\cdot|\mathcal{G})_{x=X(\omega)}$.
Hence the claim follows. 
\qed\end{proof}

\subsection{Key technical lemmas}
Let $\{\mathbb{P}_n; n=1,2,\ldots\}$ be a sequence of probability measures. The following is the key results for the current study. 
Write $\theta_\infty$ for $\{\theta_0,\theta_1,\ldots\}$. 
\begin{lemma}\label{nonrandomconv1}
Let $\theta_\infty$ be a stationary $\mathbb{P}_n$-Markov chain
with the invariant probability distribution $p_n$
that converges in probability to an ergodic Markov chain. 
Then for any bounded continuous function $\varphi$
and for any $m_n\rightarrow\infty$, 
\begin{equation}\label{eq22}
\int\varphi(\theta)p_n(d\theta)-m_n^{-1}\sum_{i=0}^{m_n-1} \varphi(\theta_i)=o_{\mathbb{P}_n}(1).
\end{equation}
\end{lemma}

\begin{proof}
Let $\mathbb{P}$ be the limit of $\mathbb{P}_n(\theta_\infty\in\cdot)$. 
Write $I^n$ and $I_{m_n}$
for the first and the
second term in the left hand side of
(\ref{eq22}) with respectively, and write $I_{i,k}$ for $k^{-1}\sum_{j=0}^{k-1}\varphi(\theta_{ik+j})$. 
Then
\begin{displaymath}
I_m=
\frac{k}{m}\sum_{i=0}^{[m/k]-1}I_{i,k}
+\frac{1}{m}\sum_{i=k[m/k]}^{m-1}\varphi(\theta_i)
\end{displaymath}
where $[x]$ is the integer part of $x\in\mathbf{R}$. 
This relation yields the upper bound of the left hand side of (\ref{eq22});
\begin{displaymath}
|I^n-I_m|\le \frac{k}{m}\sum_{i=0}^{[m/k]-1}|I^n-I_{i,k}|
+\frac{1}{m}\sum_{i=k[m/k]}^{m-1}|I^n-\varphi(\theta_i)|.
\end{displaymath}
By stationarity, each $|I^n-I_{i,k}|$ has the same law under $\mathbb{P}_n$.
Hence for a constant $C>0$ that satisfies $|\varphi(\theta)|<C$, 
\begin{displaymath}
\mathbb{E}_n[|I^n-I_m|]\le \frac{k}{m}\Big[\frac{m}{k}\Big]\mathbb{E}_n[|I^n-I_{1,k}|]+2\frac{m-k[m/k]}{m}C.
\end{displaymath}
Since $x-1<[x]\le x$ the second term is negligible
and the first term has a bound
$\mathbb{E}_n[|I^n-I_{1,k}|]$. 
Write $p$ for the limit of $p_n$ and let $I=\int\varphi(\theta)p(d\theta)$. 
Then 
$\mathbb{E}_n[|I^n-I_{1,k}|]\le \mathbb{E}_n[|I-I_{1,k}|]+|I-I_n|$
but the second term is negligible again by weak convergence of the law of the Markov chain. 
Thus the claim follows if $\mathbb{E}_n[|I-I_{1,k}|]$
can be arbitrary small. 

Since $\theta_\infty\mapsto I_{1,k}$
is continuous, 
$\mathbb{E}_n[|I-I_{1,k}|]\rightarrow \mathbb{E}[|I-I_{1,k}|]$, 
and by the law of large numbers for stationary sequence, the right hand side tends to 
$0$ as $k\rightarrow\infty$
that proves the claim. 
\qed\end{proof}

We introduce a simple sufficient condition to apply this lemma. 
Let $\mu(dx)$ be a probability measure and let $K(x,dy)$ be a transition kernel.
Let $\mu\otimes K$ be a probability measure defined by
\begin{displaymath}
(\mu\otimes K)(A\times B)=\int_AK(x,B)\mu(dx). 
\end{displaymath}
For any probability measures $p,q$ on a measurable space $(E,\mathcal{E})$  the total variation distance is
\begin{equation}\label{tv}
\|p-q\|=\sup_{A\in\mathcal{E}}|p(A)-q(A)|.
\end{equation}

%
%

\begin{lemma}\label{suflln}
Let  $K$ and $K_n\ (n=1,2,\ldots)$
be transition kernels that have the invariant probability distributions 
$p$ and $p_n$ with respectively. 
If $\|p_n\otimes K_n-p\otimes K\|\rightarrow 0$, then 
a Markov chain $\theta_\infty$ with transition kernel $K_n$ with the initial distribution $p_n$
converges in law to 
a Markov chain $\theta_\infty$ with transition kernel $K$ with the initial distribution $p$. 
\end{lemma}

\begin{proof}
It suffices to show finite dimensional convergence in law for $\theta_\infty$, and this is completed if we can prove the convergence  in total variation distance. 
Let
\begin{displaymath}
M_{m}=p\otimes  \overset{m}{\overbrace{K\otimes\cdots\otimes K}},\ M_{n,m}=p_n\otimes \overset{m}{\overbrace{K_n\otimes\cdots\otimes K_n}}.
\end{displaymath}
The task is now to show $\|M_m-M_{n,m}\|\rightarrow 0$ for any $m$. 
For $m=0,1$ the convergence is clear and assume that it is true up to $m=k$.  
For $m=k+1$, observe that $M_{k+1}-M_{n,k+1}$ equals to
\begin{displaymath}
(M_k-M_{n,k})\otimes K
+M_{n,k}\otimes (K-K_n).
\end{displaymath}
The total variation distance of the former term vanishes by assumption. For that  of the  latter,
since $p_n$ is the invariant probability distribution of $K_n$ we have 
a bound
\begin{displaymath}
\int p_n(d\theta)\|(K-K_n)(\theta,\cdot)\|\le 4\|p\otimes K-p_n\otimes K_n\|\rightarrow 0
\end{displaymath}
by Lemma 12.2.2 of \cite{L}. 
\qed\end{proof}

\subsection{Approximation of the DA procedure}\label{clc}
Let  $p_M$ be the prior distribution and 
let $P_n=\int P_\theta^{\otimes n}p_M(d\theta)$. Under some regularity conditions, 
\begin{eqnarray*}
\hat{u}=\theta+n^{-1/2}g^{-1}\eta(\theta)+o_{P_{\theta}^{\otimes n}}(1),\ 
\|p(d\theta|x_n)-N(\hat{u},n^{-1}g^{-1})\|=o_{P_{\theta}^{\otimes n}}(1)
\end{eqnarray*}
where $g=g(\hat{u})$
and $p(\cdot|x_n)$ is the posterior distribution of $M$. 
Define $P_\theta^\dagger(dy|x)$ so that $P_\theta^\dagger(dxdy)=P_\theta(dx)P_\theta^\dagger(dy|x)$.
Then under some regularity conditions, 
\begin{eqnarray*}
\hat{u}^\dagger=\theta+n^{-1/2}g^{\dagger-1}\eta^\dagger(\theta)+o_{P_{\theta}^{\dagger\otimes n}}(1),
\|p^\dagger(d\theta|x_ny_n)-N(\hat{u}^\dagger,n^{-1}g^{\dagger-1})\|=o_{P_{\theta}^{\dagger\otimes n}}(1)
\end{eqnarray*}
where 
$g^\dagger=g^\dagger(\hat{u})$ (not $g^\dagger(\hat{u}^\dagger)$)
and $p^\dagger(\cdot|x_ny_n)$ is the posterior distribution of $M^\dagger$.

Using models $M$ and $M^\dagger$, the data augmentation procedure 
is defined as the iteration of the following:
\begin{enumerate}
\item Simulate $y_n$ from $\prod_{i=1}^nP_\theta^\dagger(dy^i|x^i)=:P_\theta^\dagger(dy_n|x_n)$.
\item Simulate $\theta$ from $p^\dagger(d\theta|x_ny_n)$, 
\end{enumerate}
This procedure results in a Markov chain $\theta_0,\theta_1,\ldots$
with the invariant probability distribution $p(d\theta|x_n)$.

It is well known that this procedure is approximated by an auto-regressive process
(see \cite{SR, MengDyk99, DLR}). To explain this approximation, define 
\begin{equation}
\eta^*(\theta)
=\eta^\dagger(\theta)-\eta(\theta),\ 
g^*(\theta)
=g^\dagger(\theta)-g(\theta).
\end{equation}
Then the law of $\eta^*=\eta^*(\theta)$ tends to $N(0,g^*)$ and 
\begin{eqnarray}
\hat{u}^\dagger &=& \theta+n^{-1/2}g^{\dagger-1}(\eta+\eta^*)+o_{P_{\theta}^{\dagger\otimes n}}(1)\\
&=& \theta+n^{-1/2}g^{\dagger-1}gg^{-1}\eta+n^{-1/2}g^{\dagger-1}\eta^*+o_{P_{\theta}^{\dagger\otimes n}}(1)\\
&=& \hat{u}+g^{\dagger-1}g^*(\theta-\hat{u})+n^{-1/2}g^{\dagger-1}\eta^*+o_{P_{\theta}^{\dagger\otimes n}}(1)\label{sharp}
\end{eqnarray}
where we omit $\theta$ in $\eta(\theta)$ and $\eta^*(\theta)$.   
This calculation yields 
$n^{1/2}(\hat{u}^\dagger-\hat{u})=g^{\dagger-1}g^*\tilde{\theta}+g^{\dagger-1}\eta^*+o_{P_{\theta}^{\dagger\otimes n}}(1)
$ where $\tilde{\theta}=\sqrt{n}(\theta-\hat{u})$. 
With regularity conditions
this approximation results in a Markov chain with a transition kernel
defined by
\begin{equation}\label{Ptk}
K(\tilde{\theta},\cdot)=
N(g^{\dagger-1}g^*\tilde{\theta},g^{\dagger-1}g^*g^{\dagger-1}+ g^{\dagger-1}).
\end{equation}

\begin{remark}[Convergence Rate]
In the limit, the matrix $A:=g^{\dagger-1}g^*=I-g^{\dagger-1}g$ defines the convergence rate.  
Let $r\in [0,1)$ be the spectral radius of $A$, which is the same as the spectral radius of $g^{\dagger-1/2}g^*g^{\dagger-1/2}$. Then the marginal distribution of $\tilde{\theta}$ 
converges geometric rate $r^2$ to the invariant distribution (see Section 16.5.1 of \cite{MT}). 
A small value of $g^{\dagger-1}g$ leads to a poor performance.
\end{remark}

\begin{remark}[Invariant probability distribution]
It is easy to check the invariant probability distribution of 
$K(\tilde{\theta},\cdot)$ is $p(d\theta)=N(0, g^{-1})$ since
\begin{equation}
g^{-1}=g^{\dagger-1}g^*
g^{-1}g^*g^{\dagger-1}+
g^{\dagger-1}g^*g^{\dagger-1}+ g^{\dagger-1}.
\end{equation}
\end{remark}

\begin{remark}[Bayesian paradigm]
To show the convergence of the law of $\tilde{\theta}_0,\ldots$, 
Bayesian paradigm will be used efficiently. 
It is much difficult to show for similar method such as the stochastic EM algorithm.
This is probably the reason for the robustness of the DA procedure. 
\end{remark}

\section{Local consistency}\label{rmcp}

\subsection{Definitions of the local consistency}
\label{dmcm}

Let $(X_n,\mathcal{X}_n,P_n)\ (n=1,2, \ldots)$ be a
sequence of probability spaces.
For given $x_n\in X_n$, consider a  Markov chain 
$\theta_\infty=\{\theta_0,\theta_1,\ldots\}$ on $\mathbf{R}^d$
with the transition kernel $K$ with the invariant distribution $p$.
Note that they depends on $x_n$
and hence we will write $K(\theta,d\theta^*|x_n)$ and $p(d\theta|x_n)$. 
Write $\mathbb{P}_n$
for the joint law of $x_n$ and $\theta_\infty$. 
We call the sequence of the law $\{\mathbb{P}_n(\theta_\infty\in\cdot |x_n);x_n\in X_n\}_{n=1,2,\ldots}$
the MCMC procedure. 

\begin{definition}[Consistency]
An MCMC procedure is called consistent if 
for any $m_n\rightarrow\infty$ and for any bounded continuous function $\varphi$, 
\begin{equation}\label{consist}
\int \varphi(\theta)p(d\theta|x_n)-m_n^{-1}
\sum_{i=0}^{m_n-1}\varphi(\theta_i)=o_{\mathbb{P}_n}(1). 
\end{equation}
\end{definition}
 
The first term in the left hand side of (\ref{consist}) corresponds to the amount we are interested in 
and the second term is the Monte Carlo approximation for this amount. 
The  consistency means that this approximation tends to the targeted value after reasonable number of iteration. 
However since the posterior distribution converges to a point mass under a mild condition, 
the convergence (\ref{consist}) may not have much information. 
The  local consistency claims that the same convergence holds even after the certain scaling. 
Let $\hat{u}:X_n\rightarrow \mathbf{R}^d$
be a $\mathcal{X}_n$-measurable map
and consider
$\theta\mapsto n^{1/2}(\theta-\hat{u}).$

\begin{definition}
An MCMC procedure is called locally consistent if 
for any $m_n\rightarrow\infty$ and for any bounded continuous function $\varphi$, 
\begin{equation}\label{eq28}
\int \varphi(n^{1/2}(\theta-\hat{u}))p(d\theta|x_n)-m_n^{-1}
\sum_{i=0}^{m_n-1}\varphi(n^{1/2}(\theta_i-\hat{u}))=o_{\mathbb{P}_n}(1). 
\end{equation} 
\end{definition}

If each $\mathbb{P}_n(\theta_\infty\in\cdot |x_n)\ (x_n\in X_n, n=1,2,\ldots)$
is a stationary	process, we call the MCMC procedure stationary. 
In the main theorem in the current paper, stationarity is assumed which is an unrealistic in practice. 
The choice of the initial probability distribution is an important part for designing Monte Carlo method. 
This choice heavily depends on the structure of the model that prevents from constructing  a general framework. 
However the following illustrates that a suitable choice of the initial distribution does not change the results.
For example we can choose $N(\tilde{u}_n, n^{-1})$ as the initial distribution $q(\cdot|x_n)$ where 
$\tilde{u}_n$ is an estimator that satisfies $n^{1/2}(\tilde{u}_n-\hat{u})=O_{P_n}(1)$. 

For $\epsilon>0$, when two $\sigma$-finite measures $\mu$ and $\nu$ of $(E,\mathcal{E})$ satisfies 
$\mu(A)\le \nu(A)+\epsilon$ for any $A\in\mathcal{E}$, 
we write $\mu\le \nu+\epsilon$. 

\begin{lemma}
Consider a stationary MCMC procedure 
that has initial distribution $p(d\theta|x_n)$ which is the invariant distribution.  
Consider another MCMC procedure that replace the initial distribution 
$p(d\theta|x_n)$ to 
$q(d\theta|x_n)$. 
Then if the former MCMC procedure is consistent and if
for any $\epsilon>0$ there exists $c>0$ such that
\begin{equation}\label{nonsta}
\limsup_{n\rightarrow\infty} P_n(\{x_n;q(\cdot|x_n)> cp(\cdot|x_n)+\epsilon\})<\epsilon,
\end{equation}
the latter is also consistent. 
\end{lemma}

\begin{proof}
Let 
$A_\epsilon$ be the event that is measured by $P_n$ in (\ref{nonsta}). 
Write $\mathbb{Q}_n$ for $\mathbb{P}_n$
replacing the initial distribution from $p$ to $q$. 
For any continuous $[0,1]$-valued  function $\psi(x_n,\theta_\infty)$, we have
\begin{displaymath}
\mathbb{Q}_n(\psi)\le 
P_n(A_\epsilon)+c\mathbb{P}_n(\psi)+\epsilon.
\end{displaymath}
Hence if $\mathbb{P}_n(\psi)\rightarrow 0$, then $\mathbb{Q}_n(\psi)\rightarrow 0$. 
Take $\psi$ to be the absolute value of the left hand side of (\ref{consist}). 
\qed\end{proof}

\subsection{Local consistency of the standard the DA procedure}
Let $M$ and $M^\dagger$ be as in Section \ref{qmd} and let $p_M$ and $P_n$ be as in Section \ref{clc}. 
A sequence of tests $\psi_n:X_n\rightarrow [0,1]$
is said to be uniformly consistent 
for testing $\theta\in\mathbf{R}^d$ against $K^c\subset\mathbf{R}^d$
if 
\begin{equation}
P_\theta^{\otimes n}(\psi_n)\rightarrow 0,\ \sup_{\vartheta\in K^c}P_\vartheta^{\otimes n}(1-\psi_n)\rightarrow 0.
\end{equation}
If  there exists a uniformly consistent test
for each $\theta$ with any compact set  $K$ that includes $\theta$, $M$ is said to have a uniformly consistent test. 

Assume the following conditions. 

\begin{assumption}\label{MAass}
\begin{enumerate}
\item\label{TV1} $M^\dagger$ is quadratic mean differentiable with  same support. 
\item\label{TV2} The Fisher information matrix $g$ of $M$ is non-singular.
\item\label{TV3} $M$ has a uniformly consistent test. 
\item\label{TV4} The prior $p_M$ has a continuous, positive and bounded
density. 
\item\label{TV5} $M$ is identifiable. 
\end{enumerate}
\end{assumption}

Let 
\begin{equation}
K_n(\tilde{\theta},d\tilde{\theta}^*|x_n)=\int_{Y_n} p^\dagger(\hat{u}+n^{-1/2}d\tilde{\theta}^*|x_ny_n)P_{\hat{u}+n^{-1/2}\tilde{\theta}}^{\dagger\otimes n}(dy_n|x_n)
\end{equation}
where $\tilde{\theta}=n^{1/2}(\theta-\hat{u})$, $\tilde{\theta}^*=n^{1/2}(\theta^*-\hat{u})$.
This is the transition kernel of the DA procedure. 
The following is the main results for the current paper that says the DA procedure works well under general conditions. 

\begin{theorem}\label{regugibbs}
Assume $\theta_0\sim p(d\theta|x_n)$. Then
under Assumption \ref{MAass},
the DA procedure has the local consistency.
\end{theorem}

\begin{proof}
By  Bernstein von-Mises's theorem of the model $M^\dagger$, 
\begin{equation}\label{eq35}
\|p^\dagger(d\theta|x_ny_n)-N(\hat{u}^\dagger,n^{-1}g^{\dagger-1})\|
=o_{P_{\theta}^{\dagger\otimes n}}(1)
\end{equation}
where $g^\dagger=g^\dagger(\hat{u})$. 
It is possible to replace $\hat{u}^\dagger$ in (\ref{eq35}) by the right hand side of (\ref{sharp})
without $o_{P_\theta^{\otimes n}}(1)$ term.  
By  $\theta\mapsto \tilde{\theta}=n^{1/2}(\theta-\hat{u})$, this  is mapped to 
$g^{\dagger-1}g^*\tilde{\theta}+g^{\dagger-1}\eta^*$.
We are now in a position to show
\begin{equation}\label{eq41}
 \|\int_{y_n\in Y_n} N(g^{\dagger-1}g^*\tilde{\theta}+g^{\dagger-1}\eta^*,g^{\dagger-1})P_\theta^\dagger(dy_n|x_n)-K(\tilde{\theta},\cdot)\|=
o_{P_{\theta}^{\otimes n}}(1)
\end{equation}
for each $\theta$. 
The left hand side equals to
\begin{equation}\label{eq24}
\int_{x\in\mathbf{R}^n}
|\int\phi(x;g^{\dagger-1}\eta^*,g^{\dagger-1})
(P_\theta^\dagger(dy_n|x_n)-\phi(\eta^*;0,g^*)d\eta^*)|dx.
\end{equation}
For the moment
replace $g^\dagger$ by $g^\dagger(\theta)$
and $g^*$ by $g^*(\theta)$
and fix $x\in\mathbf{R}^d$,
Let $\psi_n:X_n\rightarrow [0,1]$ be the value in the vertical bars in (\ref{eq24}) after the replacement and set $\varphi_n\in\{-1,+1\}$ to be the sign of $\psi_n$.
Then 
for the smooth function $l(z)=\phi(x;g^{\dagger-1}(\theta)z,g^{\dagger-1}(\theta))$, 
it is sufficient to show
\begin{equation}\label{stable}
P_\theta^{\dagger\otimes n}[\varphi_nl(\eta^*)]
-
P_\theta^{\otimes n}[\varphi_n\int_{\eta^*} 
l(\eta^*)
\phi(\eta^*;0,g^*(\theta))d\eta^*]\rightarrow 0.
\end{equation}
Since $\varphi_n$ is $P_{\theta}^{\dagger\otimes n}$-tight, by choosing suitable 
probability space with a probability measure 
$P_\theta^{\dagger\otimes \infty}$, it is possible to assume $\varphi_n\rightarrow \varphi\ (n\rightarrow\infty)$. Then by replacing $\varphi_n$ by $\varphi$ in (\ref{stable}), 
the convergence follows by the stable convergence of $\eta^*$, Lemma \ref{stableconv}. 
Hence we have
\begin{equation}\label{eq26}
\|K_n(\tilde{\theta},d\tilde{\theta}^*|x_n)-K(\tilde{\theta},d\tilde{\theta}^*)\|
=o_{P_{\theta}^{\otimes n}}(1)
\end{equation}
where $K$ is defined in (\ref{Ptk}). 
By Bernstein von-Mises's theorem for model $M$, we have 
$\|p_n-p\|=o_{P_n}(1)$
where $p_n(d\tilde{\theta}|x_n)=p(\hat{u}+n^{-1/2}d\tilde{\theta}|x_n)$ and $p(d\tilde{\theta})=N(0, g^{-1})$. Hence 
 Lemma 12.2.2 of \cite{L} shows 
\begin{equation}
\|p_n\otimes K_n-p\otimes K\|
=o_{P_n}(1)
\end{equation}
since $P_\theta^{\otimes n}(dx)p_M(d\theta)=p(d\theta|x_n)P_n(dx)$
by integrating the left hand side of (\ref{eq26}) by $p_M$. 
Then Lemmas \ref{nonrandomconv1} and \ref{suflln} prove the claim
since for each $\hat{u}$, $K$ defines an ergodic Markov chain
and $\hat{u}$ is $P_n$-tight. 
\qed\end{proof}

\section{Concluding remarks}\label{cr}

\subsection{Future work}

What we did NOT discuss in this paper were the following.
We believe that the framework we proposed is helpful for these directions. 
\begin{enumerate}
\item
Research for  poor performance of the MCMC procedures. 
The local properties are helpful for identification of the performance bottlenecks of the MCMC procedures. 
This is studied in two different directions  by \cite{KKcompstat} that studies the rate of $m_n$ of (\ref{eq0})
and by \cite{Kamatani13} that studies the rate of $|\theta_i-\theta_{i-1}|$.  
\item
Research for constructing new Monte Carlo procedures.
Though the current study is for regular Monte Carlo procedures this results are useful to eliminate the  performance bottlenecks of the MCMC procedures. 
For an example, the paper \cite{Kamatani12} studies an efficient MCMC procedure for the cumulative probit model
and there are many possibilities for this direction. 
\end{enumerate}

\subsection{A technical comment}
We consider the maximum likelihood estimator $\hat{u}$. Though it does not always exist,  it can be replaced by 
the central value of the posterior distribution;
For a probability measure $\mu$ on $\mathbf{R}$, a central value is 
a point $\overline{x}\in\mathbf{R}$ satisfying 
\begin{displaymath}
\int_{\mathbf{R}} \arctan(x-\overline{x})\mu(dx)=0.
\end{displaymath}
Element of $\mathbf{R}^p$ is denoted by $x=(x^1,\ldots, x^p)^T$. 
For a probability measure $\mu$ on  $\mathbf{R}^p$, 
let $\mu^i(A)$ be $\int_{x\in\mathbf{R}} 1_A(x^i)\mu(dx)$ for $A\in\mathcal{B}(\mathbf{R})$. 
For $\mu$, 
we call
$\overline{x}=(\overline{x}^1,\overline{x}^2,\ldots,\overline{x}^p)^T\in\mathbf{R}^p$
 central value if each $\overline{x}^i$ is a central value of $\mu^i$. 
The central value always exists and  unique.
See \cite{Ito}. 


\begin{acknowledgements}
This is, essentially, the second part of the author's Ph.D. thesis
at Graduate School of Mathematical Sciences, the University of Tokyo. 
The author wishes to express his thanks to the Ph.D. supervisor, Prof. Nakahiro Yoshida for his several helpful comments and suggestions.
The author also thank to the Associate Editor and anonymous referee for constructive comments which  helped a lot to improve the paper. 
\end{acknowledgements}


\bibliographystyle{apa}      


\end{document}